\documentclass[prl,twocolumn,amssymb]{revtex4}
\usepackage{epsfig}
\def\be{\begin{equation}}
\def\ee{\end{equation}}
\def\ba{\begin{eqnarray}}
\def\ea{\end{eqnarray}}

\begin{document}
\title{Percolation of unsatisfiability in finite dimensions}
\author{J. M. Schwarz and A. Alan Middleton}
\affiliation
{Department of Physics, Syracuse University, Syracuse, New York 13244} 
\date{\today}

\begin{abstract}
The satisfiability and optimization of finite-dimensional Boolean
formulas are studied using percolation theory, rare region arguments,
and boundary effects.  In contrast with mean-field results, there is
no satisfiability transition, though there is a logical connectivity
transition. In part of the disconnected phase, rare regions lead to a
divergent running time for optimization algorithms.  The thermodynamic
ground state for the NP-hard two-dimensional maximum-satisfiability
problem is typically unique. These results have implications for the
computational study of disordered materials.
\end{abstract}
\maketitle
 
Complex problems with many degrees of freedom are of interest to both
physicists and theoretical computer scientists. The overlap is
especially strong between the physics of disordered materials and
optimization problems in the typical case.  For example, there is a
close correspondence between the ground states of Ising spin glasses,
with up and down spins, and optimal assignments of Boolean variables,
which can be true or false, in a logical formula.  This correspondence
is more than superficial as both systems exhibit phase transitions in
the structure of minimal configurations and in the dynamics of the
physical systems or optimization algorithms
\cite{BarthelWeigtHartmannMonassonZecchina}.  Such connections lead to
advances in the two fields. Combinatorial optimization algorithms from
computer science are often employed to simulate disordered condensed
matter systems \cite{HartmannRiegerBook}.  Approaches from statistical
physics, including techniques such as replica theory and concepts such
as the thermodynamic limit and scaling, have proven useful in studying
the running time algorithms and the structure of solution space
\cite{MonassonEtAl}.

Motivated by work on mean field Boolean formulas and progress in
understanding models of finite-dimensional disordered materials, we
investigate ensembles of Boolean formulas whose graphs are
two-dimensional. These formulas are composed by conjunctively joining
logical clauses, with each clause formed using nearest neighbor
variables.  The optimization problem is to assign truth values so as
to satisfy the maximum number of clauses in the formula.  This is
closely analogous to minimizing the number of broken bonds in an Ising
spin glass \cite{spinglass}. Using ideas from statistical physics,
including percolation and thermodynamic ground states, we find a
transition in the structure of logically connected components and
investigate the uniqueness of optimal assignments.

Decomposing the problem into clusters of strongly connected components
that contain contradictory cycles greatly reduces the running time of
an exact optimization algorithm.  These contradictory
strongly-connected components (CSC's) need not percolate even though
the clauses form percolative structures.  In addition, the rapid
convergence to a unique ground state as the size of the problem
increases suggests that the problem is easy in the typical case,
though it is classified as difficult in the worst case sense.  Two
central categories in this classification from computational
complexity theory are P and NP decision (yes/no) problems
\cite{Papadimitriou}.  Problems in P can be {\em decided} in time
polynomial in the size of the problem description, while a proof of
the answer for NP problems can be {\em checked} in polynomial time.
NP-hard problems, a solution to which could be used to quickly solve
any problem in NP, are believed to be solvable only in exponential
time for the worst case realizations.  It may well be that many
NP-hard problems derived from physical systems, such as finding the
ground state configuration for the 2D spin-glass in a magnetic field
\cite{Barahona}, are typically solvable in polynomial time.  Our
results support this possibility. NP-hard problems with algorithms
that typically take polynomial time on some problem sets are known
\cite{Papadimitriou}, but have not been extensively and directly
studied for physical problems in finite dimensions.

\begin{figure}[h]
\begin{center}
\epsfxsize=8.5cm
\epsfbox{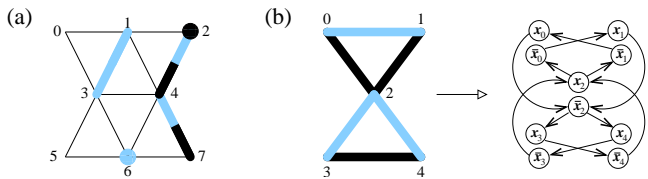}
\caption{\label{fig_Boole}(a) Example finite-dimensional Boolean
formula.  Each 2-clause in the formula is an edge, represented by two
segments.  Circles represent 1-clauses.  Black segments or circles
indicate negated variables, while the
lighter shaded segments or circles represent variables that are not negated.
The formula depicted is $(x_1\vee x_3)\wedge (x_2\vee \bar{x}_4)\wedge
(x_4\vee \bar{x}_7)\wedge (\bar{x}_2)\wedge (x_6) $.  (b) A smallest
unsatisfiable subgraph for the triangular lattice (left) and its
digraph (right).  The subgraph's formula is $(x_0\vee x_1)\wedge
(\bar{x}_0\vee \bar{x}_2)\wedge (\bar{x}_1\vee \bar{x}_2)\wedge
(x_2\vee x_3)\wedge (x_2\vee x_4)\wedge (\bar{x}_3\vee \bar{x}_4)$.  A
contradictory cycle (CC), in this digraph is $ x_2\rightarrow
\bar{x}_0\rightarrow x_1\rightarrow \bar{x}_2\rightarrow
x_3\rightarrow \bar{x}_4\rightarrow x_2$.}
\end{center}
\end{figure}

We consider finite-dimensional Boolean formulas $Z$
of the form 
\begin{equation}
Z=\wedge_{\ell=1}^M (\vee_{i=1}^{K} y_i^{\ell})
\end{equation}
where $\vee$ is the logical OR operation, $\wedge$ is the logical AND
operation, and $\{y_i^{\ell}\}$ are {\em literals} chosen from a set
$Y=\{x_1,\ldots,x_N,\bar{x}_1,\ldots,\bar{x}_N\}$ of $N$ Boolean {\em
variables} and their negations.  The variables are identified with the
vertices of a two-dimensional lattice.  We specialize to clauses with
$K=1$ and $K=2$.  We form 2-clauses by choosing two neighboring
variables and negating each variable with probability $1/2$. The
1-clauses are single literals, with probability $1/2$ of negation. A
sample formula is depicted in Fig.~\ref{fig_Boole}(a). The ensemble is
defined by parameters $\alpha$ and $\gamma$, respectively the ratios
of the number of $2$-clauses to $N$ and $1$-clauses to $N$.  The
2-clauses do not overlap and no two 1-clauses contain the same
variable.  Given a truth assignment $x_i\rightarrow \{T,F\}$ for all
Boolean variables, a clause is satisfied if one of the literals in the
clause is $T$.  If all clauses are satisfied, the formula $Z$ is
satisfied.  Determining the existence of a satisfying truth assignment
is the problem of satisfiability (SAT).

The optimization of the number of satisfied clauses in $Z$ can be
mapped to determining the ground state of a spin glass in a
heterogeneous field.  This mapping translates Boolean assignments
$x_i=\{\rm{F,T}\}$ to spin variables $S_i=\{-1,+1\}$.  A bond energy
$E_{\ell}$ can be assigned to a clause $(y_0^\ell\vee y_1^\ell)$
connecting variables $x_i$ and $x_j$ via \cite{MonassonEtAl}
\begin{eqnarray}
E_\ell=\frac{J}{4}[1-\Delta(y^\ell_0)S_i - \Delta(y^\ell_1)S_j 
+ \Delta(y^\ell_0)\Delta(y^\ell_1)S_i S_j],
\end{eqnarray}
where $\Delta(y^\ell_0)=1$ if $y^\ell_0=x_i$ and $\Delta(y^\ell_0)=-1$
if $y^\ell_0=\bar{x}_i$ and similarly for $j$ (nearest neighbor to
$i$) replacing 0 with 1.  The total spin glass energy $E$ is given by
$E=\sum_{\ell=1}^M E_\ell$.  Any clause that is not satisfied costs an
energy $J$; the existence of an $E=0$ ground state is equivalent to
satisfiability of the Boolean formula.

Resolution \cite{Papadimitriou} is a method that can be used to
quickly decide SAT for $K\le 2$. This procedure is equivalent to
mapping each 2-clause to a pair of logical implications and searching for
``contradictory cycles'' (CCs).  For example, the clause $x_1 \vee
x_2$ is equivalent to $\bar{x}_1\rightarrow x_2$ and
$\bar{x}_2\rightarrow x_1$. Clauses with $K=1$ are replaced by a
single implication, e.g., $\bar{x}_1$ becomes $x_1 \rightarrow
\bar{x}_1$.  The Boolean formula can be represented by an implication
digraph (i.e., directed graph) $G=(Y,E)$ with $2N$ vertices and $(2\alpha +
\gamma) N$ edges $E$. For a sample mapping, see
Fig. \ref{fig_Boole}(b).  The formula $Z$ can not be satisfied if
there is a CC, which is a path $p$ in $G$ that connects a variable to
its negation and vice versa, i.e., $p=(x_i\rightarrow x_j \ldots
\rightarrow \bar{x}_i \rightarrow \ldots \rightarrow x_i)$.  For the
formulas we consider, the existence of contradictory cycles can be
decided in time linear in $N$ \cite{Cormenetal}.

\begin{figure}[h]
\begin{center}
\epsfxsize=8cm
\epsfbox{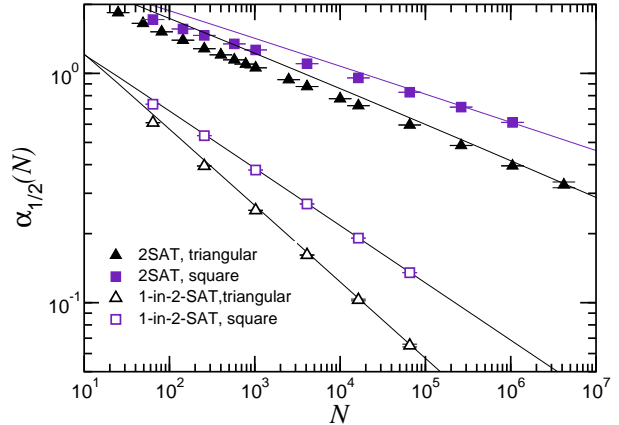}
\caption{\label{fig_coarse}(a) Plot of $\alpha_{1/2}(N)$, the clause
density at which $1/2$ of the graphs are satisfiable, as a function of
lattice size $N$.  Symbols indicate numerical results for 2SAT and
1-in-2-SAT (Ising spin glass) on triangular and square
lattices. Curves are analytic approximations found in a small subgraph
expansion.  }
\end{center}
\end{figure}

We find that there are CCs for any $\alpha>0$ (taking $\gamma=0$), as
$N\rightarrow\infty$, in these finite-dimensional formulas.  Defining
$\alpha_{1/2}(N)$ as the value of $\alpha$ for which $1/2$ of the
finite-dimensional $N$-variable formulas are satisfiable,
$\alpha_{1/2}(N)\rightarrow 0$ as $N\rightarrow\infty$ (see
Fig.~\ref{fig_coarse}.)  This crossover is coarse, in that the width
of the crossover from low to high probability of satisfiability is
proportional to $\alpha_{1/2}(N)$, for large $N$.  This to be
contrasted with random mean-field $K=2$ formulas, where for
$N\rightarrow\infty$, there is a sharp SAT to UNSAT phase transition (the
probability that a formula is satisfiable is $1$ for
$\alpha<\alpha_c=1$ and $0$ for larger $\alpha$.)  These differences
result from small CCs, which at small $\alpha$ are exponentially rare
in the mean field case but appear with Poissonian statistics in the
finite-dimensional case, where loops are more important.

The location of the SAT/UNSAT crossover can be computed by an
expansion in $\alpha$. Some subgraphs are ``forcing'', i.e., in all
satisfying assignments one of the variables has a fixed truth value.
The smallest unsatisfiable graph is found by joining two contradictory
forcing subgraphs.  An example of this graph type is depicted in
Fig.~\ref{fig_Boole}(b). On the triangular lattice, these subgraphs
have density $\rho_{\triangle}(\alpha)=\frac{\alpha^6}{2^7
3^4}+\mathcal{O}(\alpha^7)$.  The density of the simplest
unsatisfiable graphs on the square lattice is $\rho_{\Box}(\alpha)
=\frac{\alpha^8}{2^{16}}+\mathcal{O}(\alpha^9)$.  In general, if the
smallest unsatisfiable subgraph has $r$ bonds and density
$c_r\alpha^r$, the probability of satisfiability is $P_{\rm
SAT}(N)=(1-c_r\alpha^r)^N$, to lowest order in $\alpha$, giving
$\alpha_{1/2}(N)\approx(c_r^{-1/r}\ln 2)N^{-1/r}$.  We plot numerical
results and analytic expansions for $\alpha_{1/2}(N)$ in
Fig.~\ref{fig_coarse}, which includes the next order analytic
corrections in $\alpha$ (7-edged subgraphs with density
$\frac{5^2}{2^7 3^7}\alpha^7+{O}(\alpha^{8})$ on the triangular
lattice and 9-edged subgraphs with density
$\frac{\alpha^9}{2^{16}}+\mathcal{O}(\alpha^{10})$ on the square
lattice).

We also plot analytic estimates and numerical results for
$\alpha_{1/2}(N)$ for the 1-in-2-SAT problem in
Fig.~\ref{fig_coarse}. While a clause in 2SAT (i.e., $K=2$)
is satisfied if either
literal is true, a clause is satisfied in 1-in-2-SAT when exactly one
literal in a clause is true.  The 1-in-2-SAT problem maps both to an
Ising spin glass in the absence of a magnetic field and to the
two-color problem \cite{Walsh}. The smallest unsatisfiable graphs are
given by frustrated cycles, giving $\alpha_{1/2}(N)\approx 3(N/\ln
2)^{-1/3}$ and $\alpha_{1/2}(N)\approx 2^{5/4}(N/\ln 2)^{-1/4}$ for
the triangular and square lattices, respectively.

Given the lack of a sharp SAT/UNSAT transition, due to the existence
of {\em small} unsatisfiable graphs, we have investigated the
percolation of {\em large} unsatisfiable graphs as a phase
transition. We study these graphs within the context of MAXSAT, which
is the problem of minimizing the number of unsatisfied clauses.  In
two dimensions, the determination of the ground state for the Ising
spin glass (or MAX-1-in-2-SAT) is in P \cite{Barahona}, while
determining the ground state for MAXSAT with $K=2$ is
NP-hard, even for planar graphs.
We studied the CSCs,
sets of literals for which any two literals are connected by a
directed path in the implication digraph and which contain a CC.  We
find that the probability of having a spanning CSC has a transition
that becomes sharper with increasing $N$, with a critical value for
$\alpha$ of $\alpha_S=1.8245(5)$ on the triangular lattice.
The cluster size distribution $n(s)$ at criticality behaves as
$n(s)\sim s^{-\tau}$, with $\tau=2.02(5)$. The scaling of the
probability for a spanning CSC near $\alpha_S$ gives a correlation
length exponent of $\nu=1.32(3)$.  These values are consistent with
the 2D values for standard percolation, where $\tau=187/91$ and
$\nu=4/3$ \cite{StaufferAharony}.

Percolation of paths in the implication digraph can be related to
connectivity percolation \cite{Bollobasetal}. This is done by
projecting part of the implication digraph on $2N$ literals onto an
undirected graph of $N$ variables.  This ``trimmed'' projection has
the same statistics as standard connectivity percolation with edge
probability $\tilde{p}=2p-p^2$, where for our finite-dimensional case,
$p=\alpha/2z$, with $z=4$ ($z=6$) for square (triangular) lattices, if
overlapping clauses are allowed. (On the triangular lattice with
overlapping clauses, we find $\alpha_S=1.887(2)$.)  In mean field, the
percolation of paths containing a contradiction coincides with the
SAT/UNSAT transition. This is not the case in finite dimensions. But
given this type of connection and studies of percolation of directed
edges in finite dimensions \cite{PfeifferRieger}, it is not surprising
that CSC percolation appears to be in the same universality class as
standard connectivity percolation.

\begin{figure}[h]
\begin{center}
\epsfxsize=8cm
\epsfbox{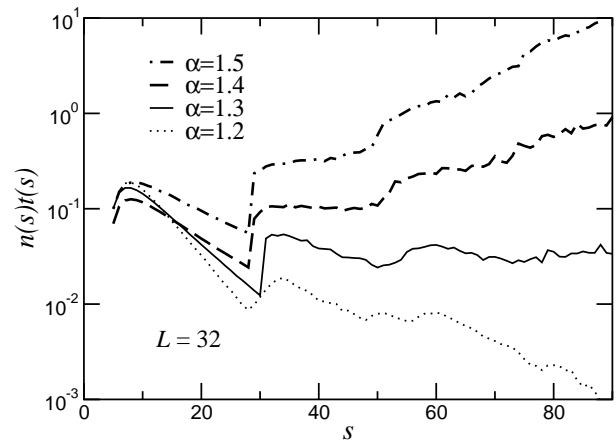}
\caption{\label{fig_timing}Convolution of the median number of
backtracks $t^*(s)$ with the CSC cluster size distribution $n(s)$,
where $s$ is the cluster mass.  Note that $\alpha<\alpha_S$ for these
curves.}
\end{center}
\end{figure}

The decomposition of the graphs into CSCs speeds up exact search
algorithms for MAXSAT.  Here, we apply this decomposition to estimate
running times of such an algorithm. We used a MAXSAT code \cite{bf}
that first finds a heuristic bound to the solution and then applies an
exact Davis-Putnam-Loveland-Logemann (DPLL) search.  The running time
measure $t$ is the number of ``backtracks'' that are executed while
partially exploring the tree of all possible assignments.  Each CSC
cluster can be loaded into the algorithm individually
\cite{footindep}.  The sum of the unsatisfied clauses from each
cluster gives the minimal number of unsatisfied clauses for the
entire formula.  When $\alpha<\alpha_S$, the distribution of sizes of
the CSCs, is exponentially decaying in the cluster size, $n(s)\sim
e^{-s/s_\xi(\alpha)}$, with $s_\xi \propto
\xi^d\propto(\alpha_S-\alpha)^{-d \nu}$.  If we plot the median number
of backtracks for each cluster, we find that the median running time
of the DPLL-type algorithm scales exponentially with the cluster size,
${t^*}(s)\sim e^{s/s_\tau(\alpha)}$.  When $s_\xi(\alpha) <
s_\tau(\alpha)$, the median running time for a sample, $T^*(L)$, is
bounded by a multiple of the system volume, $T^{*}(L)\sim L^2$.
However, when $s_\xi > s_\tau$, $T^*(L)$ diverges more rapidly, with
an estimate for the largest cluster size in a finite sample giving
$T^*(L) \sim L^{2s_\tau/s_\xi}$. The mean running time, ${\overline
T}(L)$, diverges exponentially with $L$.  The transition between the
linear and superlinear median time behaviors defines
$\alpha_G<\alpha_S$ via $s_\xi(\alpha_G)=s_\tau(\alpha_G)$.
Fig.~\ref{fig_timing} shows convolutions of the cluster size
distribution $n(s)$ and the median time $t^*(s,\alpha)$ as a function
of size. The change from negative to positive slope on the semi-log
plot gives $\alpha_G\approx 1.3$ for the DPLL code we use.  This
slowing down of the algorithmic dynamics is similar to that for the
physical dynamics of random magnets \cite{GriffithsRanderia} and is
reminiscent of the change from the easy-SAT to hard-SAT phases in
random graphs \cite{BarthelWeigtHartmannMonassonZecchina}.

Despite the divergence of the running times for the exhaustive
DPLL-type algorithms, we might expect that the ground states could be
found in time proportional to the system volume in the typical case,
even above the CSC percolation transition.  Assuming that the droplet
picture describes these finite-dimensional spin glasses, the presence
of the magnetic field destroys the spin glass phase \cite{FisherHuse}
and the correlations are finite-ranged (though in 2SAT there are some
correlations in the external fields.)  So while the CSCs percolate,
the effects of frustration remain localized beyond some length scale.
This picture also implies one unique thermodynamic state.  If this is
the case there may be a way to develop a new algorithm to deal with
the local frustrated bonds, either by solving subsystems and
joining the solutions together to form the whole system or by a more
clever heuristic algorithm.  We leave this as a conjecture and simply
test for uniqueness.
  
To test whether the ground state is unique, we study the effect of
boundary conditions, similar to studies of the Ising spin glass
\cite{NSISGunique}.  By comparing ground states for a system of linear
size $L$ and an expanded system of linear size $L'$ (each with free
boundaries), one can determine if the ground state is unique or not
from the sensitivity to boundary conditions.  If the solutions in the
common subsystem of linear size $w$ become fixed as $L$ and $L'$
diverge, a unique ground state exists in the thermodynamic limit. Note
that the ground state must be unique for $\alpha < \alpha_S$, as the
logical structure of the graph does not percolate.

Since the $\pm J$ spin glass with magnetic field (equivalent to
optimal assignments for MAX2SAT) has many degenerate ground states, we
study the weighted MAX2SAT (WMAXSAT) question, where the degeneracy is
broken, to be able directly to compare ground state solutions. Each
clause has an associated weight, chosen uniformly in the interval
$[0,1)$, and the optimization problem is now to minimize the sum of
the weights of the unsatisfied clauses.  We also introduce 1-clauses,
with the same weight distribution. The addition of 1-clauses lowers
$\alpha_S$, allowing us to study a larger range system of system
sizes, as the graphs are sparser.

\begin{figure}[h]
\begin{center}
\epsfxsize=8cm
\epsfbox{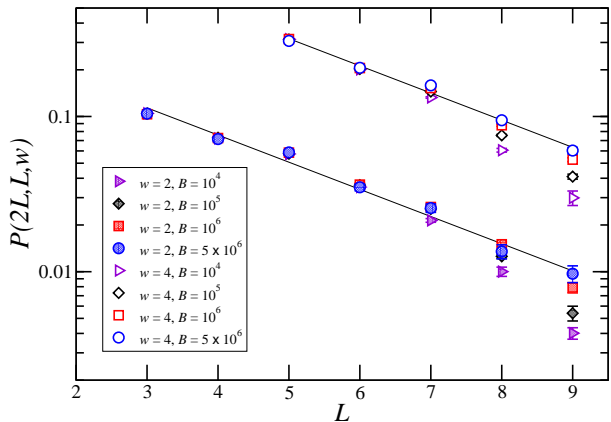}
\caption{\label{fig_unique}Log-linear plot of $P(2L,L,w)$ for
$\alpha=1.7$ and $\gamma=0.2$ for weighted MAX2SAT.  The bound on backtracks is
$B$.  The lines are exponential fits for the
$w=2,4$, $B=5\times 10^6$ data.}
\end{center}
\end{figure}

We estimate the quantity $P(2L,L,w)$, the probability that there is a
change in the central area of area $w^2$ when the system size expands
from $L$ to $2L$ \cite{NSISGunique}, by sampling from the WMAXSAT
ensemble.  To be able to complete the simulations, we impose an upper
limit $B$ on the number of backtracks in the DPLL and report $P$ for a
range of $B$. The points in Fig.~\ref{fig_unique}, with $w=2$, are
well fit by an exponential in $L$, in the limit of large $B$. (A power
law fit gives an exponent less than $-2$, which is inconsistent with a
fractal domain wall picture for a model with two states
\cite{NSISGunique}.)  The $w=4$ data is also well described by an
exponential with the same slope. For $\alpha=1.7$ and $\gamma=0.2$, we
estimate a correlation length of $\xi = 2.5\pm 0.3$. The exponential
approach to a unique state holds for all $\alpha$ and $\gamma$ that we
explored.

Working within the concepts and the algorithms for spin glasses and
other disordered materials, we have studied the problem of optimal
satisfaction of Boolean formulas.  There is no thermodynamic SAT to
UNSAT transition, due to the finite density of small unsatisfiable
formulas.  There is a percolation transition, however, in the logical
structure of the formulas as the clause density is increased, that is
apparently in the class of standard connectivity percolation. Below
this transition, we use rare region arguments to predict a transition
in the mean running time of an optimization algorithm.  We find that
the ground state is unique even in the high clause density
regime. This uniqueness suggests that the MAX2SAT problem can be
solved ``locally'' by studying subsamples larger than the correlation
length and patching subsolutions together (though for large
correlation lengths, rare regions might again dominate the running
time.)  This general approach in turn has potential applications to
algorithms for studying spin glasses and other random magnets.  This
project was supported in part by the National Science Foundation 
through grant DMR-0109164.


\begin{thebibliography}{99}
\bibitem{footindep} As each CSC is strongly connected,
assignment of truth values to
variables outside of a given CSC does not affect
satisfying assignments for that CSC. Note that weighted
MAXSAT might lead to CSC-CSC interactions.
\end{thebibliography}
\end{document}